\documentclass[11pt,a4paper]{article}
\usepackage{lipsum}
\usepackage{authblk}
\usepackage{amsmath,a4}  
\usepackage[utf8]{inputenc}
\usepackage{graphicx,enumerate,color,amsfonts,amsthm,amsmath,amssymb,mathrsfs,marginnote,dsfont,textcomp,tikz,marginnote,subfigure,pdflscape,multirow,booktabs,float,enumitem,scrtime,currfile,titleps,hyphenat,relsize,psfrag}
\usepackage[colorlinks,
    linkcolor={red!60!black},
    citecolor={black},
    urlcolor={blue!80!black}]{hyperref}
\usepackage{tikz,pgf,pgffor,pgfplots}
\tikzset{>=latex}
\usetikzlibrary{calc,patterns,decorations,intersections,decorations.pathmorphing,fit,backgrounds,arrows,shapes,automata,petri,positioning}
\usepgfmodule{shapes,plot}

\theoremstyle{definition}

\theoremstyle{remark}

\makeatletter
\newcommand{\setword}[2]{%
  \phantomsection
  #1\def\@currentlabel{\unexpanded{#1}}\label{#2}%
}
\makeatother

\allowdisplaybreaks
\usepackage[top=2cm, bottom=2cm, left=2cm, right=2cm]{geometry}
\usepackage{fancyhdr}
\newpagestyle{main}{
\setheadrule{.4pt}
\sethead{}
        {}
        {\sectiontitle}
}

\renewenvironment{abstract}{%
\hfill\begin{minipage}{0.95\textwidth}
\rule{\textwidth}{1pt}}
{\par\noindent\rule{\textwidth}{1pt}\end{minipage}}
\makeatletter
\renewcommand\@maketitle{%
\hfill
\begin{minipage}{0.95\textwidth}
\vskip 2em
\let\footnote\thanks 
{\Large \bf \@title \par }
\vskip 1.5em
{\large \@author \par}
\end{minipage}
\vskip 1em \par
}
\makeatother

\begin{document}
%
\title{\nohyphens{A modeling study of predator--prey interaction propounding honest signals and cues$^*$}}
\author[1]{Ahd Mahmoud Al-Salman}
\author[2,1]{Joseph P\'aez Ch\'avez}
\author[1,$\ast$]{Karunia Putra Wijaya}

\affil[1]{\small\emph{Mathematical Institute, University of Koblenz, 56070 Koblenz, Germany}}
\affil[2]{\small\emph{Center for Applied Dynamical Systems and Computational Methods (CADSCOM), Faculty
of Natural Sciences and Mathematics, Escuela Superior Polit\'ecnica
del Litoral, P.O. Box 09-01-5863, Guayaquil, Ecuador}}

\affil[$\ast$]{Corresponding author. Email: \href{mailto:karuniaputra@uni-koblenz.de}{karuniaputra@uni-koblenz.de}}
\maketitle
\begin{abstract}
Honest signals and cues have been observed as part of interspecific
and intraspecific communication among animals. Recent theories
suggest that existing signaling systems have evolved through natural
selection imposed by predators. Honest signaling in the
interspecific communication can provide insight into the evolution
of anti-predation techniques. In this work, we introduce a
deterministic three-stage, two-species predator--prey model, which
modulates the impact of honest signals and cues on the interacting
populations. The model is built from a set of first principles
originated from signaling and social learning theory in which the
response of predators to transmitted honest signals or cues is
determined. The predators then use the signals to decide whether to
pursue the attack or save their energy for an easier catch. Other
members from the prey population that are not familiar with
signaling their fitness observe and learn the technique. Our
numerical bifurcation analysis indicates that increasing the
predator's search rate and the corresponding assimilation efficiency
gives a journey from predator--prey abundance and scarcity, a stable
transient cycle between persistence and near-extinction, a
homoclinic orbit pointing towards extinction, and ultimately, a
quasi-periodic orbit. A similar discovery is met under the increment
of the prey's intrinsic birth rate and carrying capacity. When both
parameters are of sufficiently large magnitudes, the separator
between honest signal and cue takes the similar journey from a
stable equilibrium to a quasi-periodic orbit as it increases. In the
context of modeling, we conclude that under prey abundance,
transmitting error-free honest signals leads to not only a stable
but also more predictable predator--prey dynamics.
~
\end{abstract}

\section{Introduction}

Biological signal has been one of the fundamental subjects in
understanding interactions among animals. Owren \textit{et al.}
\cite{owren2010redefining} suggest that a biological signal carries
motivational messages, which contain signaler arousal and likely
upcoming actions. Smith \cite{smith1997behavior} and Hauser
\cite{hauser1996evolution} characterize a biological signal as
information sharing, which is a way broader than carrying solely
emotional-motivational messages. Bugnyar \textit{et al.}
\cite{bugnyar2001food} give a relying \cite{owren2010redefining} yet
less popular definition of a biological signal, that is,
\emph{encoded information} about sender's features and stimuli.
Lehmann \textit{et al.} \cite{lehmann2014cues} mention that a
biological signal represents evolved communication that influences
the receiver behavior throughout efficiently altering sender's
information, which in turn affects the cooperative tendencies of the
receiver. However, the optimality of signaling for a signaler
depends on the receiver's interpretation, meaning that responses
count on the signal \cite{johnstone1992continuous}. The continuing
evolution of biological signals forces signalers and receivers to be
engaged in a battle of extracting correct interpretation of the
signals. It is not immediately apparent whether or not signals are
honest or dishonest, making individuals more conspicuous
\cite{leal1999honest}.

Zahavi in \cite{zahavi1975mate,zahavi1977cost,zahavi1987theory}
argues that biological signals are always honest, and they aim to
show off the hidden quality of the signaller to the receiver.
Dawkins and Krebs \cite{dawkins1978animal,krebs1984behavioral}
oppose Zahavi's idealism, claiming that organisms bluff and cheat in
their signals when communicating with each other. \emph{Zahavi's
handicap hypothesis}
\cite{zahavi1999handicap,grafen1990biological,zahavi1975mate} is
explained particularly for defending honest and costly signals
against such counter-arguments. The essence of the hypothesis is
that an individual's quality must ``honestly'' be signaled to others
because signaling is costly. However, Sz\'{a}mand\'{o} and Penn
\cite{szamado2015does} disagree with this hypothesis for several
reasons, mainly because there are versions of the handicap
hypothesis, but none of which succeeded in providing theoretical
proofs. Although a game-theoretic Evolutionarily Stable Strategy
(ESS) model \cite{grafen1990biological}, which models the handicap
principle in sexual selection, can be excepted from that conclusion,
other authors
\cite{bergstrom2002separating,getty1998handicap,getty2006sexually,lachmann2001cost,szamado2011cost}
did challenge the model. In \cite{polnaszek2014not}, the authors
provided a first experimental conclusion regarding honesty. Their
laboratory experiment using blue jays (Cyanocitta cristata) showed
that signal's cost stabilizes thus encourages honesty, which in
return confirms the handicap hypothesis. The model in
\cite{clifton2016handicap} describes biological signals as mostly
honest, at least when there is a large enough variance in
individual's quality. In the realm of animal communication, costly
signals are entailed with the consistency of the population and
increased predation risk \cite{zahavi1975mate}. However, for a
signal to be an ESS, its costs should decrease the net gain of
signaling in order to maintain the signal reliability
\cite{hasson1997towards}. Johnstone and Grafen in
\cite{johnstone1992error} argue that despite the importance of the
handicap principle in biological communication, it comes with an
unrealistic assumption that existing supporting models assume
error-free biological communications. A general error-prone
evolutionarily stable signaling model of the handicap principle is
presented by Johnstone and Grafen in \cite{johnstone1992error}, that
incorporates a perceptual error. A numerical analysis of such
perceptual error in the model is presented in
\cite{johnstone1994honest}. The bottom line is, regardless of
whether the signaling systems are error-prone or error-free, they
must be honest at equilibrium \cite{johnstone1992error}.

\paragraph{Honest signal.}
The study of biological signals reveals information about the
signaller and the receiver, where in the context of the current
study, prey and predators. In fact, honest signals elevate the
ability of prey to escape an attempt of predation
\cite{zahavi1975mate,zahavi1987theory}. In other words, honest
signals are adaptive behavior of prey to help them convey their
alertness and relative escape ability from a predator's attack.
Predators then decide whether to use the information to attack or
not. Predators may as well assess such individual quality, due to
which case the honest signaling continues to evolve. An example of
honest signal is \emph{stotting} in Thomson's gazelle, which is
acted through jumping highly near to a predator instead of running
away. This helps signal gazelle's fitness and give a clear message
to the predator that ``you cannot catch me'' \cite{smith2003animal}.
Among biological signals such as acoustic signals, where animals
show themselves by laud vocalization, and aposematic signals, when
revealing conspicuous colors and even repeated movements, the
freedom from error (honesty) is determined by how useful the
information is for the sender and the receiver
\cite{laidre2013animal}. A fundamental lack of symmetries in signals
communication is derived from the different roles and interests of
the signaller and receiver, highlighting the importance of
explication study on influential signals rather than informative
signals \cite{rendall2009animal}.

\paragraph{Cues.}
The behavioral change of animals in conveying information about
themselves is significant for both the signaler and the receiver.
The behavioral change is considered to be positive when the signaler
conveys honest information about its fitness, which in turn benefits
both the signaler and receiver. In contrast, it is considered to be
negative when the information benefits the receiver exclusively. In
studies of animal behavior, a fundamental distinction is created
between signals. On the one side is honest signal, which is a
perceivable entity that has evolved for a specific purpose of
conveying information to alter the receiver's behavior, and on the
other side is cue, which is unintentional information disclosed by
the signaler benefitting the receiver exclusively. An example of a
cue is the rustling sound of a mouse that runs through the
underbrush. A predator can apparently use this sound as information
that may lead it to the mouse's location. However, this information
is primarily produced by the mouse rustling activity rather than
developed by learning to transmit information.

The primary objective of this study is to model the impact of honest
signals and cues on the sustenance of predator--prey interactions.
We propose a two-species predator-prey model with the prey
population split into individuals who are capable of transmitting
honest signals (experienced) and those who are not (inexperienced).
At this point, we declare the assumption that all the transmitted
signals from experienced prey are honest. However, these signals are
\emph{error-prone}, meaning that they are subject to different
interpretations from distinct predator individuals. Signal
transmitting is deemed as a behavioral trait that occurs between
experienced and inexperienced prey until the latter attain an
experience level and emerge into the veteran population. Such
process affects the predator, experienced and inexperienced prey
population, albeit in different ways. One of the primary concerns of
this study is to describe these effects on the population level.

\section{Mathematical model}

Our primary goal in this section is to build a system of
differential equations that govern the population dynamics of a
predator-prey system in which experienced prey individuals share
their skills in transmitting honest signals with inexperienced prey.
Experienced prey individuals are knowledgeable and well-informed,
comprising those that are capable of transmitting honest signals
against attacks. In other words, they can transmit honest signals to
a predator to increase their survival chances. On the other hand,
inexperienced prey individuals cannot transmit honest signals but
discern signals transmitted by the experienced prey and learn how to
evolve their escaping techniques. In the system, the prey population
that is constrained only by limited environmental resources, is the
only source of nourishment for the predator. Throughout this work,
we denote by $x$, $y$ and $z$ the density of inexperienced prey,
experienced prey and predator individuals, respectively.

\subsection{Prey population growth in the absence of predators}

In realistic ecosystems, the population growth rate cannot be
represented without environmental constraints. It may grow
exponentially for some time, but they will ultimately curtail owing
to ceilings determined by limited resources availability. Since
every newborn prey is inexperienced, the exponential growth can be
modeled by $r(x+y)$, where $r$ denotes the intrinsic growth rate.
The growth term $r(x+y)$ assumes that both inexperienced and
experienced prey are able to reproduce. The competition for
logistics can usually be modeled by $(r/C)x(x+y)$, where $C$ denotes
the carrying capacity, i.e., the maximal prey density the
environment can support. The term $x(x+y)$ represents the
mass-action law accounting for competition or interaction for
seizing logistics among $x$--individuals and between $x$-- and
$y$--individuals. Consequently, the logistic growth model of the
prey population in the absence of predator is represented by the
following equations
\begin{equation}\label{eq:prey1}
\begin{split}
x'&=r(x+y)-\frac{r}{C}x(x+y),\\
y'&=-\frac{r}{C}y(x+y).
\end{split}
\end{equation}
Observe that the entire prey population density $x+y$ satisfies the
standard logistic equation.

\subsection{Predation effect on the prey population}

One of the most important components to interpret a mathematical
model of the predator--prey system is functional response. In this
study, a predator-dependent functional response is considered where
prey abundance changes because of predators' interference. Let
$\beta$ denote the predator's search rate (sometimes referred to as
\emph{scanning speed}), determining how much area is covered by one
predator per unit time. During a predetermined search time $T_s$,
the predator is able to cover the area $\beta T_s$. Following
Holling's work \cite{holling1959components}, we impose an assumption
that on every event of meeting, a prey individual can always be
captured by the predator. Thus, the total prey individuals of
$X_c:=\beta T_s x$ are successfully captured by one average predator
during $T_s$. When the aforementioned assumption is dropped, Holling
uses an inhibition factor $k$ such that $X_c:=\beta T_s x^k$
instead, which later sets his functional response of type III. Let
$\tau$ denote the time an average predator needs to ``handle'' one
prey individual (handling time). It is worth mentioning that $\tau$
also includes the satiation time, food-sharing time, and time of not
motivated to hunt (cf. \cite{jeschke2002predator}). Then, $T_h:=\tau
X_c$ represents the total time needed by one predator to handle the
entire captured prey individuals. When $T:=T_s+T_h$ denotes the
total time for a predator to search and handle the entire captured
prey individuals, we get Holling's type II functional response
$X_c=\beta Tx \slash (1+\beta\tau x)$, which measures the total
number of captured prey individuals by one predator during $T$. We
note that in general, $\tau\ll T$. As no conflict among predators
during hunting is assumed, the total of $[\beta Tx \slash
(1+\beta\tau x)]\cdot z$ prey individuals per unit area can
successfully be captured by $z$ predators during the entire time
$T$, or $[\beta x \slash (1+\beta\tau x)]\cdot z$ individuals per
unit area and unit time. Including predation, the $x$--dynamics in
\eqref{eq:prey1} is corrected as
\begin{equation}\label{eq:prey2}
x'=r(x+y)-\frac{r}{C}x(x+y)-\frac{\beta x}{1+\beta \tau x}\cdot z.
\end{equation}
At this point, the prey dynamics can be set down by combining the
logistic growth with the functional response. The latter is
proportional to the prey density but saturates to a certain maximum
due to an intrinsic capacity of the predators.

\subsection{Cue for risk assessment by experienced prey}

Similarly as for the inexperienced prey, predation leads to decrease
in the experienced prey population, notwithstanding the fact that
the functional responses cannot be exactly the same. Although the
experienced prey can actively convey information and influence the
behaviour of the predator through transmitting honest signals, these
signals can still be subject to flaws such as leaving different
interpretations from some predator individuals at different ranges.
A possibility that an error-prone honest signal may turn into a cue
is thus envisaged. Rather than using an inhibition factor like in
Holling's type III functional response, this study engineers the
parameter $\rho\in [0,1]$ indicating the presence of error-prone
honest signals:
\begin{equation}\label{eq:prey4}
y'=-\frac{r}{C}y(x+y)-\rho\cdot \frac{\beta y}{1+\beta\tau y}\cdot
z.
\end{equation}
Actual cues are passive, as they unintentionally provide the
receiver with information that might be correlated with a high
chance of attacking \cite{lehmann2014cues}. Many studies in the
literature refer cues to as non-evolving biological traits because
of their impact on the species \cite{smith1995animal}. Honest
signals are also subject to turning into cues as they highly count
on receiver's interpretation \cite{johnstone1992continuous}. In
Thomson's gazelles, for example, actual cues attract predators to
come and any honest signal (stotting) performed in less than 40
meters away from predators becomes meaningless \cite{FF1988}. This
happens as if the prey individual continuously produces cues. When
$\rho$ is small, an average honest signal contains enough
information for the predator to rely on such that the catch is
undone \cite{laidre2013animal}. This is when the signal is entirely
\emph{error-free}. However, in a blunder or conflicting interest,
the maintenance of reliable signals is up to the evolutionary
interest of the predator, and therefore, signals become
uninformative and turn into cues. That is when the signals contain
error and $\rho$ is close to $1$.

\subsection{Learning boosting the cast for experienced prey}

Many studies found that animals can recognize their predators
without prior experience. It means that predator recognition is a
strong genetic component \cite{csanyi1993learning}. Nevertheless,
genetically functioned defense patterns could be modified by
learning \cite{csanyi1993learning}. This response modification that
is relative to the experienced prey can be seen as an adaptive
mechanism, which is a result of \emph{social learning}. Social
learning occurs when individuals adapt to new behavior or
information regarding their environment by observing and interacting
with other species \cite{brown2003social}. The authors in
\cite{boyd1988culture} demonstrate that locally adaptive ecological
behavior from experienced individuals assumes to be beneficial. In
our context, experienced prey $y$ and predator density $z$ play an
important role in this regard, in which case we assume that the
\emph{learning rate} $L=L(y,z)$ appear with those arguments. The
more intakes or attacks the predator do, the more signals the
experienced prey transmit. In return, this will decrease the
learning duration $L^{-1}$.

We shall mention certain assumptions to reveal an explicit formula
for $L$. We assume that the encounter between experienced prey and
predators enhances learning, therefore $L\sim yz$. A constant value
for $L$ is assigned when the predator density $z$ is known to
achieve a certain threshold, say $\eta$. Without loss of generality,
we assume that $L=1$ when $z=\eta$. The proposed biological scenario
thus designates the unit time (time scale) in our model as the
inexperienced prey's learning duration when $z=\eta$, which could be
several days. The learning rate must then decrease or increase, i.e.
$L<1$ or $L>1$, when $z$ turns to be smaller or larger than the
threshold $\eta$. This study considers a situation where the
interspecific competition between predators and prey is so intense
that an average experienced prey individual has multiple encounters
with predators in one day. A number of studies have shown that
maternal distress in some vertebrates caused by predator encounters
affect offspring's brain, which leads to attention and learning
deficits \cite{Wein2008,Scho2011,Macca2014}. General anxiety,
depression, and less physical activeness are among other conspicuous
marginal effects. Here, we take into account some proportion of the
inexperienced prey (typically infants and juveniles) being products
of such intense encounters. Prominent from learning deficit and less
activeness is how these individuals heavily rely on parental or
alloparental guard against predator's attacks. We thus impose an
assumption based on this reliance that the learning rate $L$
decreases only slowly when more parents and alloparents are present
but predators are scarce, i.e. when $y$ increases and $z<\eta$, and
increases only slowly when $y$ increases but the predators are
overwhelming $z>\eta$. The latter stems from high urgency in
mastering the escaping technique, adding external trigger to
learning. Based on the previous descriptions, we take the following
ansatz from several possibilities in forming rational functions due
to its small number of parameters used:
\begin{equation}\label{eq:learning}
L(y,z):=\frac{\alpha+yz}{\alpha+\eta y}.
\end{equation}
We assume that both $\alpha$ and $\eta$ are positive. Since there is
no clear indicator whether a certain age is deemed ready for
learning, we immerse the age transition term for the inexperienced
prey to the learning rate. Realistically, the carrying capacity $C$
is defined in such a away that $x,y\leq C$, making $L(y,z)\geq
L(C,0)>0$ for all nonnegative $y,z$ and that learning and age
transition are unstoppable natural components of prey population.

\subsection{Predator population growth model}

In this study, we assume that there is no other food resources to
maintain predators' survival except the prey. This means in the
absence of prey, predators are subject to starvation and their
numbers are expected to decline exponentially
\begin{equation*}
z'=-\theta z.
\end{equation*}%
In the presence of prey, however, this decline is opposed by birth
of predators as a function of prey abundance. In the classical
predator--prey models, the number of newborns is usually assumed to
be proportional to the biomass resulted from predation
\begin{equation}\label{eq:predator}
z'=\delta\cdot\frac{\beta x}{1+ \beta \tau x}\cdot
z+\delta\cdot\rho\cdot\frac{\beta y}{1+ \beta \tau y}\cdot z-\theta
z.
\end{equation}
Here, $\delta$ represents an assimilation efficiency, i.e. the
efficiency of predator in converting prey biomass into offsprings.

\section{Predator--prey model highlighting honest signals, cues and learning}

\subsection{Model summary and scaling}

The ideas of the previous discussions can now be assembled to form a
set of three differential equations that describe the biological
dynamics of a two-species predator-prey model
\begin{equation} \label{model}
\begin{split}
x' &= r(x+y)-\frac{r}{C}x(x+y)-\frac{\beta x}{1+\beta \tau x}\cdot z -\frac{\alpha+yz}{\alpha+\eta y}\cdot x, \\
y' &=\frac{\alpha+yz}{\alpha+\eta y}\cdot x - \frac{r}{C}y(x+y)-\rho\cdot \frac{\beta y}{1+\beta\tau y}\cdot z,\\
z' &=\delta\cdot\frac{\beta x}{1+ \beta \tau x}\cdot
z+\delta\cdot\rho\cdot\frac{\beta y}{1+ \beta \tau y}\cdot z-\theta
z.
\end{split}
\end{equation}
Having a system of eleven parameters, we will first consider
suitable parameters and variable transformations in order to reduce
the model complexity, without affecting the main biological
features. Let us introduce the scaling parameters $X,Y,Z$ and
dimensionless variables based on the scaling, $\tilde{x}:=x\slash
X$, $\tilde{y}:=y\slash Y$, and $\tilde{z}:=x\slash Z$. We also
foresee by the formulation of the learning rate that the parameter
$\alpha$ might not be measurable, i.e. unobservable. Additional to
the time scaling, we suppose that $X,Y,Z$ and $\alpha$ are chosen in
such a way that $\sqrt{\alpha}=\eta=X=Y=Z$. Therefore, the central
of the scaling is the observable (specifiable) threshold $\eta$. We
acquire a modification \eqref{model} as follows
\begin{equation*}
\begin{split}
\tilde{x}'&= r\left(\tilde{x}+\tilde{y} \right) -\frac
{rX}{C}\tilde{x} \left( \tilde{x}+\tilde{y} \right)-
{\frac {\beta X\tilde{x}}{1+\beta X\tau\tilde{x}}}\cdot \tilde{z}-\frac { \left( \alpha+X^2\tilde{y}\tilde{z}
\right)}{\alpha+X^2\tilde{y}}\cdot \tilde{x}, \\
\tilde{y}'&= \frac { \left( \alpha+X^2\tilde{y}\tilde{z}
\right)}{\alpha+X^2\tilde{y}}\cdot \tilde{x} -
\frac {rX}{C}\tilde{y} \left( \tilde{x}+\tilde{y} \right)-\rho\cdot{\frac { \beta X\tilde{y}}{1+\beta X\tau\tilde{y}}}
\cdot \tilde{z}, \\
\tilde{z}'&=\delta\cdot{\frac{ \beta X\tilde{x}}{1+\beta X\tau
\tilde{x}}}\cdot \tilde{z}+\delta\cdot\rho\cdot{\frac{\beta
X\tilde{y}}{1+\beta X\tau\tilde{y}}}\cdot \tilde{z}-\theta\tilde{z}.
\end{split}
\end{equation*}
To avoid hulking notation in the above system, we backtrack to the
original variables and parameters only for simplicity. In other
words, $x\leftarrow \tilde{x}$, $y\leftarrow \tilde{y}$,
$z\leftarrow \tilde{z}$. Additional to this, we suppose that
$\beta\leftarrow \beta X$ and $K:= rX\slash C$. The resulting
simplified model which we primarily focus in the sequel can now be
read as follows
\begin{equation} \label{eq:scaled}
\begin{split}
x' &= r(x+y)-Kx(x+y)-\frac{\beta x}{1+\beta \tau x}\cdot z -\frac{1+yz}{1+ y}\cdot x, \\
y' &=\frac{1+yz}{1+ y}\cdot x - Ky(x+y)-\rho\cdot \frac{\beta y}{1+\beta \tau y}\cdot z,\\
z' &=\delta\cdot\frac{\beta x}{1+ \beta\tau x}\cdot
z+\delta\cdot\rho\cdot\frac{\beta y}{1+ \beta\tau y}\cdot z-\theta
z.
\end{split}
\end{equation}

\subsection{Parameter ranges}

Here, we assume that $r\in [0,1]$ and $C\gg 1$. Ideally, the
threshold for the predator density $\eta$ to actuate the intensive
learning holds $\eta\ll C$, facilitating the abundance of prey.
Therefore, we come into the specification $K\in [0,1]$. The
unobservability of the search rate (scanning speed) $\beta$ also
hints that the new $\beta\leftarrow \beta X$ lives in
$\mathbb{R}_+$, which depends on the type of predator considered.
Recall from the model that $\tau\ll T$, from which when the time
scale and $T$ are comparable, we can assume $\tau\in [0,1]$. We also
have as previously discussed that $\rho\in [0,1]$. Certain
predators, e.g. snakes, can lay eggs more than the number of prey
individuals (e.g. frogs) eaten in one time scale. Others like lions
produce much fewer offspring as compared to the number of captured
prey individuals. This study considers the later type of predator
where $\delta\in [0,1]$. Finally, the predator lifespan duration
$\theta^{-1}$ is assumed to be greater than the learning duration of
inexperienced prey when the threshold condition holds $z=\eta$,
therefore $\theta\in [0,1]$.

\subsection{Equilibria}

Let us suppose that all the parameter values in the
system~\eqref{eq:scaled} are positive. The system exhibits two
analytically presentable equilibria:
\begin{alignat*}{2}
E_1&=(0,0,0)\quad &&\text{(\emph{extinction equilibrium})},\\
E_2&=(x_2,y_2,0)\quad &&\text{(\emph{predator extinction
equilibrium})},
\end{alignat*}%
where
$$x_2 :=(y_2r+r)y_2\quad\text{and}\quad y_2 := \frac{\sqrt{(K+Kr)^2+4Kr^2}-(K+Kr)}{2Kr}.$$
For further use and brevity, let us denote by
$$(f,g,h),\,w\in\{\beta,\delta,K,r,\rho\},\, \arg_{w}\ell (w)$$
the vector field in the system~\eqref{eq:scaled}, a representative
of the chosen bifurcation parameters, and positive solution(s) $w$
such that $\ell(w)=0$, respectively. In the forthcoming study of
codim--1 bifurcations, we assume to have fixed other parameter
values once $w$ is chosen.

The first finding regarding  $E_1$ reveals that the Jacobian matrix
$\nabla_{(x,y,z)}(f,g,h)|_{E_1}$ has the eigenvalues $-1,r,-\theta$.
This means that the system has a one-dimensional unstable manifold
and a two-dimensional stable manifold around $E_{1}$. In other
words, $E_1$ has the configuration of a saddle node. Moreover, any
value of $w$ forming a concatenated point $(w,E_1)$ cannot be a
bifurcation point (branching point) due to non-singularity of the
Jacobian matrix, see Crandall and Rabinowitz \cite{Crandall1980}.

As far as $E_2$ is concerned, the following results are expected.
The block matrix $\nabla_{(x,y)}(f,g)|_{E_2}$ has eigenvalues
$-r,\,-2r\sqrt{(K+Kr)^2+4Kr^2}\slash(\sqrt{(K+Kr)^2+4Kr^2}-K+Kr)$.
Furthermore, the full Jacobian matrix has zero entries on the third
row except on the diagonal, whose entry is given by
\begin{equation}\label{eq:eig}
\lambda_0:=\frac{\delta \beta x_2}{1+\beta\tau
x_2}+\frac{\delta\rho\beta y_2}{1+\beta\tau y_2}-\theta.
\end{equation}
This reveals that the model~\eqref{eq:scaled} has a stable manifold
around $E_2$ of dimension at least $2$. We also know that any value
of $w$ concatenated with $E_2$ cannot form a branching point except
$(\arg_w\lambda_0,E_2)$, providing $\arg_w\lambda_0$ exists, because
the Jacobian matrix is singular at the point. Let us suppose that
certain numerical values for all parameters except $w$ are given
such that $\arg_w\lambda_0$ exists. Our next aim is to confirm that
$(\arg_w\lambda_0,E_2)$ is indeed a branching point and a continuum
of nontrivial (coexistence) equilibria bifurcates from the point
under perturbation of $w$ around $\arg_w\lambda_0$. We argue that
direct substitutions and solving a higher-order polynomial of either
of the three state variables returns complications, in which case we
favor the aid of bifurcation theory. The analysis centers on the
behavior of the equilibrium due to perturbation of $w$ when the
equilibrium is sufficiently close to $E_2$.

The vector field $(f,g)$ generates a system of algebraic equations
where $x$ and $y$ change as a response of perturbation of $z$. We
know that in some subset of $\mathbb{R}^3$ where $x,y$ are
nonnegative, the vector field $(f,g)$ is $C^{\infty}$. According to
Implicit Function Theorem, the nonsingularity of the block matrix
$\nabla_{(x,y)}(f,g)|_{E_2}$ gives neighborhoods $U(x_2,y_2)$ of
$(x,y)$ and $V(0)$ of $z$ where $(x,y)=(x(z),y(z))\in
C^{\infty}(V(0))$ solves the equilibrium equation $(f,g)=0$ for any
given $z\in V(0)$ and $\lim_{z\rightarrow 0}(x,y)=(x_2,y_2)$ in the
neighborhood $U(x_2,y_2)$. For further use, let us define
$$
M:=\left\lVert \nabla^{-1}_{(x,y)}(f,g)\cdot \left(\begin{array}{c}
\partial_zf\\
\partial_zg
\end{array}
\right)\right\rVert_{L^{\infty}(U(x_2,y_2),V(0))}
$$
that specifies a Lipschitz constant of $x(z)$ and $y(z)$, i.e.
\begin{equation}\label{eq:bound}
|x-x_2|\leq M |z|\quad\text{and}\quad |y-y_2|\leq M|z|
\end{equation}
for all $z\in V(0)$. The Implicit Function Theorem helps guarantee
that $M$, which contains the evaluated right-hand sides of $x'(z)$
and $y'(z)$ on the neighborhoods, exists and is bounded. The yet
untreated equilibrium equation $h=0$ from \eqref{eq:scaled} can now
be rearranged via Taylor expansions of $\beta x\slash (1+\beta \tau
x)$ around $x_2$ and $\beta y\slash (1+\beta \tau y)$ around $y_2$
and \eqref{eq:bound} into the \emph{canonical form} of bifurcation
equation
\begin{equation}\label{eq:canonic}
h(z)=\lambda_0z-\lambda z+\omega(z)=0, \,\text{where
}\lambda=0,\,\omega(z)=\mathcal{O}\left(|z|^2\right) \text{ as
}z\rightarrow 0
\end{equation}
and $\lambda_0$ is as given in \eqref{eq:eig}. Apparently, the
equation \eqref{eq:canonic} resembles the Lyapunov--Schmidt reduced
form \cite{Rabi1977,Golu1985}. When a small positive $z$ is shown to
solve \eqref{eq:canonic}, the continuity of $x(z)$ and $y(z)$
guarantees the existence of a coexistence equilibrium. Our next aim
is thus to synthesize conditions from the canonical
form~\eqref{eq:canonic} under which a positive $z$ bifurcates from
$0$. Work on analyzing such canonical form has been ubiquitous,
which typically includes Brouwer Degree Theory, Brouwer Index
Theory, and Rabinowitz's Global Bifurcation Theorem. However, it is
not our aim in this paper to recall the exhaustive content of the
theories. Despite instantly picking up corresponding terminologies,
we will place the accompanying references for further reading,
whenever necessary.

Consider the canonical form~\eqref{eq:canonic} where $h:Q\rightarrow
\mathbb{R}$ for some open interval $Q\subset \mathbb{R}$ and $h\in
C^{\infty}(Q)$. A point $s\in \mathbb{R}$ where $s\notin h(\partial
Q)$ is called \emph{regular} if either $h^{-1}(s)=\emptyset$ or all
points $z^{\ast}\in h^{-1}(s)$ make $h'(z^{\ast})$ invertible,
otherwise it is called \emph{critical}. The map
$\mathcal{B}:C^1(Q)\times Q\times \mathbb{R}\rightarrow \mathbb{Z}$
defined as
$$
\mathcal{B}(h,Q,s):=\begin{cases}
\sum_{z^{\ast}\in h^{-1}(s)}\text{sign}(h'(z^{\ast})), & s\text{ regular},\\
\mathcal{B}(h,Q,\tilde{s}),& s\text{ critical, }\tilde{s}\text{
regular, with }\lVert s-\tilde{s}\rVert<\inf_{m\in h(\partial
Q)}\lVert s-m\rVert,
\end{cases}
$$
denotes the \emph{Brouwer degree} of $h$ in an interval $Q$ with
respect to a reference point $s$ \cite{Ma2005,Krasno1984}. When
$z=0$ is an isolated singular value of $h$ (i.e. no further singular
value of $h$ in a certain neighborhood $Q(0)$ of $0$), then the map
$$
\mathcal{I}(h,0):=\mathcal{B}(h,Q(0),0)
$$
defines the \emph{index} of $h$ at the isolated singular value
$z=0$, which is derived from Brouwer degree of $h$ in a neighborhood
of $z=0$ with respect to the reference point $s=0$. At this stage,
we know that whether $z=0$ is isolated or not depends on the
neighbourhood $Q(0)$ taken. Confirming that $(\arg_w\lambda_0,0)$ is
a branch point of \eqref{eq:canonic} requires that the index changes
in value when $w$ is perturbed around $\arg_w\lambda_0$
\cite{Ma2005}. Let us thus suppose that $0=\lambda\neq \lambda_0$
such that the polynomial $h$ in \eqref{eq:canonic} has a nonzero
slope at $z=0$ (i.e. $h'(0)=\lambda_0-\lambda\neq 0$) and choose
$Q(0)$ sufficiently small such that $h^{-1}(0)=\{0\}$. The former
indicates that $z=0$ is an isolated singular value and $s=0$ is
regular. The index is then given by
$$
\mathcal{I}(h,0)=\text{sign}(h'(0))=\text{sign}(\lambda_0-\lambda)=
\begin{cases}
1,& \lambda_0>\lambda=0\\
-1,& \lambda_0<\lambda=0
\end{cases}.
$$
This claims that $(\arg_w\lambda_0,0)$ is indeed a branching point
of \eqref{eq:canonic}. Let us suppose that
$$w>\arg_w\lambda_0\Leftrightarrow \lambda_0>0.$$
One can pick an example to this case for suitable numerical values
when $\lambda_0$ changes from negative to positive as $\beta$
increases. Suppose that $\lambda_0\neq \lambda=0$, further
rearrangement of \eqref{eq:canonic} into
$$
z-\lambda\mathcal{L}z+\tilde{\omega}(z)=0,\,\text{where
}\mathcal{L}:=\lambda_0^{-1},\,\tilde{\omega}(z)=\mathcal{L}\omega(z)=\mathcal{O}\left(|z|^2\right)
\text{ as }z\rightarrow 0,
$$
fits into the framework in \cite[pp. 161--166]{Cushing1988} and
\cite{Wijaya2017}, where the eigenvector corresponding to the largest eigenvalue of $\mathcal{L}$ gives the
leading order (initial direction) of the continuum of bifurcating
positive solutions, according to the representation
\begin{equation*}
z=\text{eigvec}\{\mathcal{L}\}\cdot \varepsilon +
\mathcal{O}(\varepsilon^2)
\end{equation*}
for a sufficiently small $\varepsilon>0$. In our current study, the
condition
\begin{equation}
w>\arg_w\lambda_0
\end{equation}
shall give $\mathcal{L}=\lambda_0^{-1}>0$. According to the
aforementioned references, a continuum of positive solutions $z$ to
\eqref{eq:canonic} bifurcates from $(\arg_w\lambda_0,0)$ with the
note that $w$ is sufficiently close to $\arg_w\lambda_0$.

As the reminder, we note that the preceding exposition solely scopes
out the behavior of equilibrium around a branching point and the
existence of a branching continuum of coexistence equilibria around
the point, excluding the uniqueness. To go beyond both cases, we
perform a numerical bifurcation analysis employing all the
aforementioned bifurcation parameters in the next section.

\section{Numerical study of the predator-prey model via path-following methods}

In this section, our main concern will be to investigate the
dynamical response of the predator-prey system~\eqref{eq:scaled} as
certain model parameters are varied. Specifically, we will take the
predator's search rate $\beta$, the predator's assimilation
efficiency $\delta$, the intrinsic growth rate of prey $r$, the
competition coefficient $K$ and the separator $\rho$ as our main
bifurcation parameters. The numerical investigation will be carried
out using the path-following software COCO (short form of
Computational Continuation Core \cite{dankowicz2013}). This is an
analysis and development platform for the numerical treatment of
continuation problems using MATLAB. A remarkable feature of COCO is
its set of toolboxes that covers, to a good extent, the
functionality of available continuation packages, such as AUTO
\cite{auto97} and MATCONT \cite{matcont}. In particular, in this
section we will make extensive use of the COCO-toolboxes `ep' and
`po', which encompass a set of routines for the bifurcation analysis
of parameter-dependent families of equilibria and limit cycles,
respectively, in smooth dynamical systems.
\begin{table}[H]
    \centering
    \renewcommand\arraystretch{1.2}
    \begin{tabular}{|c | l | l | c | c |}
        \hline
        Parameter  & Description & Unit & Range & Reference value\\ \hline
        $r$ & Prey's intrinsic growth rate  & $[\text{t}^{-1}]$ & 0--1 & 0.8 \\
        $K$ & Competition coefficient & $[\text{t}^{-1}]$ & 0--1 & 0.1\\
        $\beta$ & Scaled predator's search rate & $[\text{a}\times \text{ind} \times \text{t}^{-1}]$ & 0--3 & 1.7\\
        $\tau$ & Predator's handling time & $[\text{t}]$ & 0--1 & 0.3\\
        $\rho$ & Separator between honest signal and cue & -- & 0--1 & 0.5\\
        $\delta$ & Predator's assimilation efficiency & -- & 0--1& 0.7\\
        $\theta$ & Predator's natural death rate & $[\text{t}^{-1}]$ & 0--1 & 0.1 \\ \hline
    \end{tabular}
    \caption{\label{tab:cont}Parameter values used for the numerical bifurcation analysis. The unit symbol t, a, ind
     stand for unit time, unit area, and individual, respectively.}
\end{table}

\begin{figure}[H]
\centering
\psfrag{a}{\large(a)}\psfrag{b}{\large(b)}\psfrag{c}{\large(c)}
\psfrag{B}{\Large$\beta$}\psfrag{d}{\Large$\delta$}\psfrag{z}{\Large$z$}\psfrag{x}{\Large$x$}
\includegraphics[width=\textwidth]{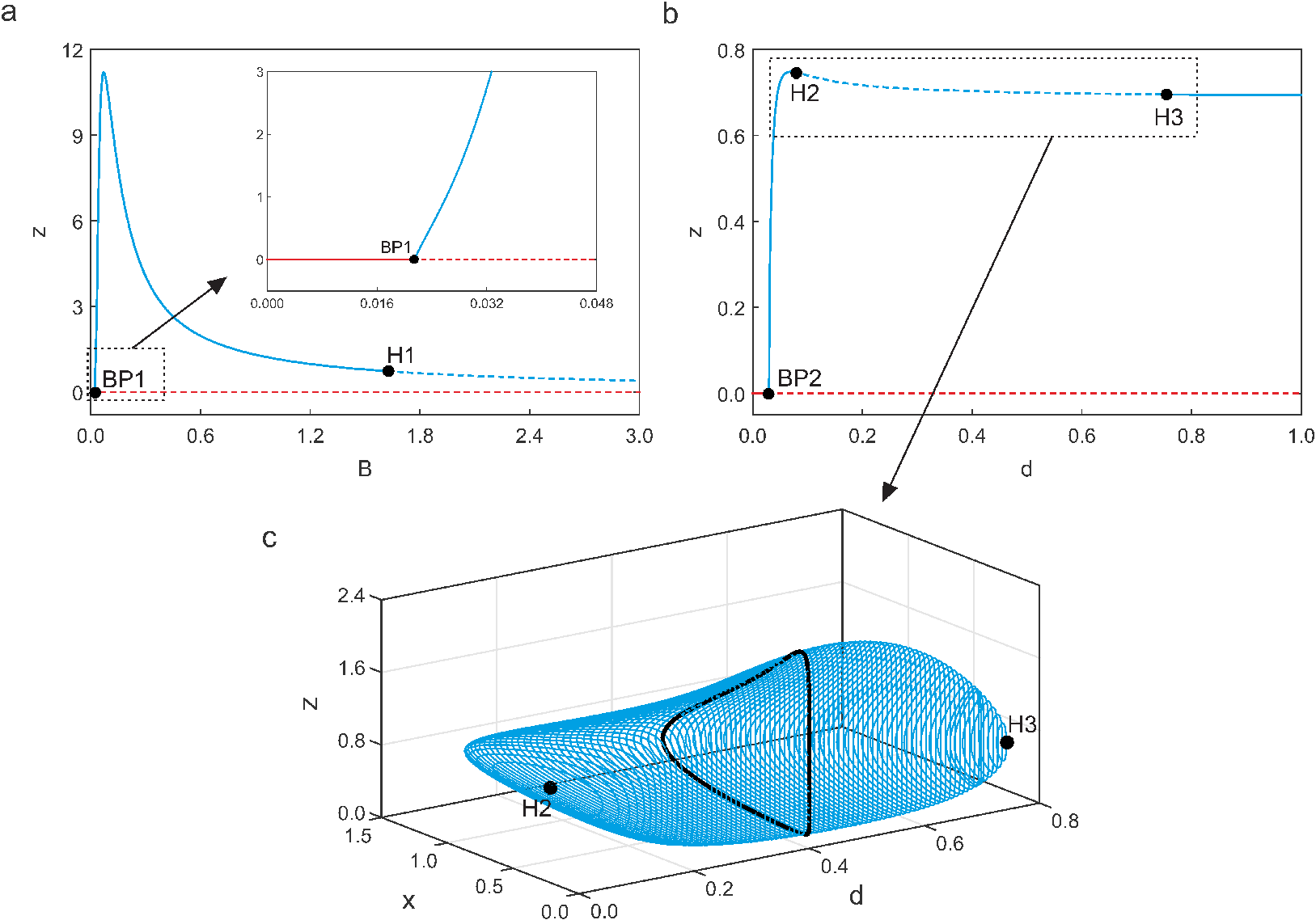}
\caption{One-parameter continuation of equilibria of
system~\eqref{eq:scaled} with respect to the predator's search rate
$\beta$ (panel (a)) and the predator's assimilation efficiency
$\delta$ (panel (b)), computed for the parameter values given in
Table~\ref{tab:cont}. In these panels, the blue and red curves
represent continuation of endemic and predator-free equilibria,
respectively. Solid and dashed lines depict stable and unstable
solutions, respectively. Bifurcation points are highlighted as
follows: BP$i$ (branching points) and H$i$ (Hopf points). Panel (c)
shows the solution manifold obtained via continuation of periodic
solutions between the Hopf points H2 and H3 found in panel (b). The
figure shows a cross section corresponding to a periodic solution at
$\delta=0.4$.}\label{Figure1}
\end{figure}

Let us begin our study by analyzing the behavior of the equilibria
of the model~\eqref{eq:scaled} as the predator's search rate $\beta$
and the predator's assimilation efficiency $\delta$ vary, one at the
time. The result can be observed in Fig.\ \ref{Figure1}.
In the figure, the blue and red curves represent continuation of
coexistence and predator extinction equilibria, respectively. Panel
(a) reveals that for small values of $\beta$ the system possesses a
stable predator extinction equilibrium, meaning that under this
condition any initial predator population will decay in time. As
$\beta$ increases, a critical value
$\beta=\arg_{\beta}\lambda_0\approx0.02144$ (labeled BP1) is
detected, where the predator extinction equilibrium loses stability.
From this point on, another branch (in blue) emerges, corresponding
to the numerical continuation of stable coexistence equilibria,
where the predator population can now be sustained. This abrupt
change in the system behavior corresponds to a branching point (of
supercritical type) \cite{guck,govaertslibro}, also known in the
Mathematical Biology field as a forward bifurcation
\cite{velasco2000,watmough2000}. For $\beta$ larger than this
branching point, a maximal value of the predator population size
(under equilibrium) is found at $\beta\approx0.07049$. This value
can be interpreted as the optimal predator's search rate, in the
sense that smaller $\beta$ means the predator population is not
taking advantage of the available food source provided by the prey,
while $\beta$ too high means a faster decrease of the prey
population, making the predator's food source scarce. Another
critical point is found at $\beta\approx1.62824$, where a
supercritical Hopf bifurcation H1 takes place, hence giving rise to
a family of stable limit cycles, while the equilibrium becomes
unstable. A similar scenario occurs when the predator assimilation
efficiency $\delta$ is varied, see Fig.\ \ref{Figure1}(b).
In this case, as before, small $\delta$ makes the predator
population unsustainable in the system, until the critical value
$\delta=\arg_{\delta}\lambda_0\approx0.02963$ is reached, where the
system undergoes another forward bifurcation (BP2). For larger
values, the bifurcation analysis reveals the presence of a family of
limit cylces, limited by two Hopf points (H2 at
$\delta\approx0.07964$, H3 at $\delta\approx0.75458$), hence
defining a window of sustained oscillations with amplitudes strongly
dependent on the predator's assimilation efficiency $\delta$. Thus,
the system presents a branch of stable oscillatory behavior forming
a closed loop, starting and ending at the Hopf points H2 and H3, see
Fig.\ \ref{Figure1}(c). This dynamical phenomenon is
referred often to as \emph{endemic bubble} \cite{liz2015}.

\begin{figure}[H]
\centering \psfrag{a}{\large(a)}\psfrag{b}{\large(b)}
\psfrag{B}{\Large$\beta$}\psfrag{y}{\Large$y$}\psfrag{x}{\Large$x$}
\psfrag{Zavg}{\Large$z_{\mbox{\scriptsize mean}}$}
\includegraphics[width=\textwidth]{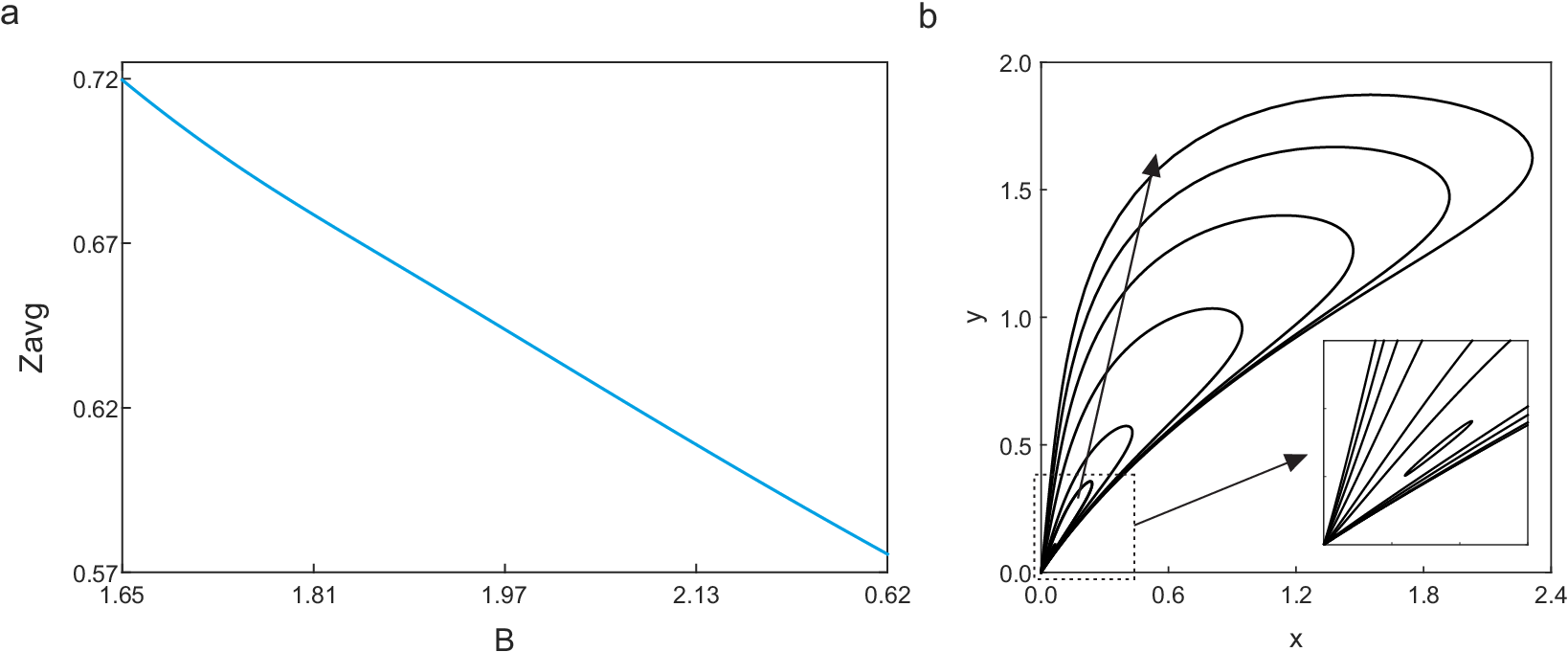}
\caption{(a) One-parameter continuation of periodic solutions of
system~\eqref{eq:scaled} with respect to the predator's search rate
$\beta$, computed for the parameter values given in
Table~\ref{tab:cont}. In this panel, the vertical axis shows the
mean value of the $z$-component of the periodic response. Panel (b)
depicts a sequence of closed orbits obtained along the curve shown
in panel (a). Here, the direction of increasing $\beta$ is indicated
by an arrow.}\label{Figure2}
\end{figure}

\begin{figure}[H]
\centering
\psfrag{a}{\large(a)}\psfrag{b}{\large(b)}\psfrag{c}{\large(c)}\psfrag{d}{\large(d)}\psfrag{e}{\large(e)}
\psfrag{B}{\Large$\beta$}\psfrag{dd}{\Large$\delta$}\psfrag{x}{\Large$x$}\psfrag{y}{\Large$y$}
\psfrag{z}{\Large$z$}\psfrag{t}{\Large$t$}
\includegraphics[width=\textwidth]{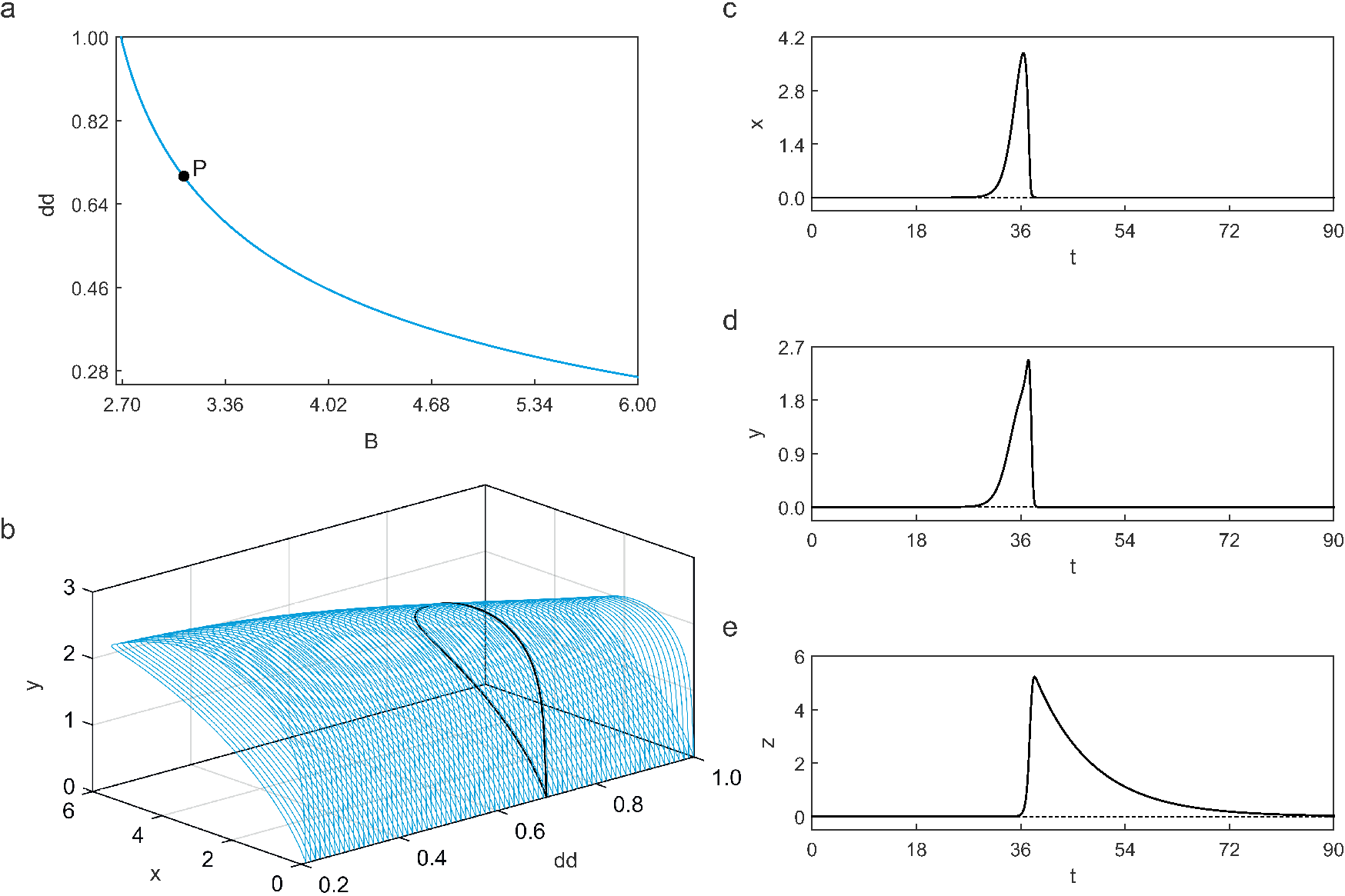}
\caption{(a) Two-parameter continuation of a family of periodic
solutions with high period approximating a homoclinic orbit, with
respect to $\beta$ and $\delta$, for the parameter values given in
Table~\ref{tab:cont}. Panels (c)--(e) present time plots of a
solution computed at the test point P ($\beta=3.0918$,
$\delta=0.7$). Panel (b) depicts the solution manifold computed
along the curve in panel (a), showing a cross section corresponding
to the test solution found at P.}\label{Figure3}
\end{figure}

Next, we will analyze in more detail the periodic response of the
predator-prey model considered here. For this purpose, we will fix
all parameter values according to Table~\ref{tab:cont} and let the
predator search rate $\beta$ vary for $\beta>1.62824$, that is,
after the Hopf bifurcation at H1 occurs. The result of the numerical
continuation of the parameter-dependent family of periodic solutions
is shown in Fig.\ \ref{Figure2}. Panel (a) shows the
behavior of the mean value of the $z$-component of the periodic
response as $\beta$ varies. Panel (b) depicts the family of periodic
orbits on the $x$-$y$ plane. This figure suggests that the sequence
of limit cycles converges to a homoclinic solution, i.e., a
trajectory that approaches a saddle equilibrium for $t\to\pm\infty$.
Typical state-space representations of near-homoclinic orbits are
characterized by the development of a ``corner'' in the phase-plot
located close to the saddle-equilibrium, as can be detected in Fig.\
\ref{Figure2}(b), while the period of the limit cycles
increases showing long excursions around the equilibrium (in this
case the extinction equilibrium $x=y=z=0$) due to the slow dynamics.
Biologically speaking, this means that the predator and prey
populations present fluctuations with episodes of almost-extinction
of long duration. The oscillatory behavior indicates that the
ecosystem provides suitable conditions for all populations to
recover again after the near-extinction phase. At the homoclinic
solution, however, the three populations can experience only one
phase of growing behavior after which all populations go extinct. In
other words, only one generation can be sustained by the ecosystem,
see Fig.\ \ref{Figure3}(c)--(e).

As is well known, a family of near-homoclinic orbits can be computed
via numerical continuation of periodic solutions with large (fixed)
period \cite{dankowicz2013}, as a homoclinic solution can be found
as the limit of a branch of periodic orbits when the period tends to
infinity, connecting a saddle equilibrium with itself. In this way,
we will compute a branch of near-homoclinic orbits when the predator
search rate $\beta$ and the predator assimilation efficiency
$\delta$ vary simultaneously, see Fig.\
\ref{Figure3}(a). Thus, we obtain an approximation of
a family of $(\beta,\delta)$-values where periodic solutions
disappear via a homoclinic bifurcation. Panel (b) depicts the
solution manifold projected on the $x$-$y$ plane along the
near-homoclinic curve. Here, it can be seen that all solutions have
the extinction equilibrium as the base point. Panels (c)--(e)
present the time profiles of all populations showing the rise and
decay phases described in the previous paragraph.

\begin{figure}[H]
\centering
\psfrag{a}{\large(a)}\psfrag{b}{\large(b)}\psfrag{c}{\large(c)}\psfrag{d}{\large(d)}\psfrag{e}{\large(e)}
\psfrag{f}{\large(f)}
\psfrag{B}{\Large$\beta$}\psfrag{dd}{\Large$\delta$}\psfrag{z}{\Large$z$}\psfrag{t}{\Large$t$}
\includegraphics[width=\textwidth]{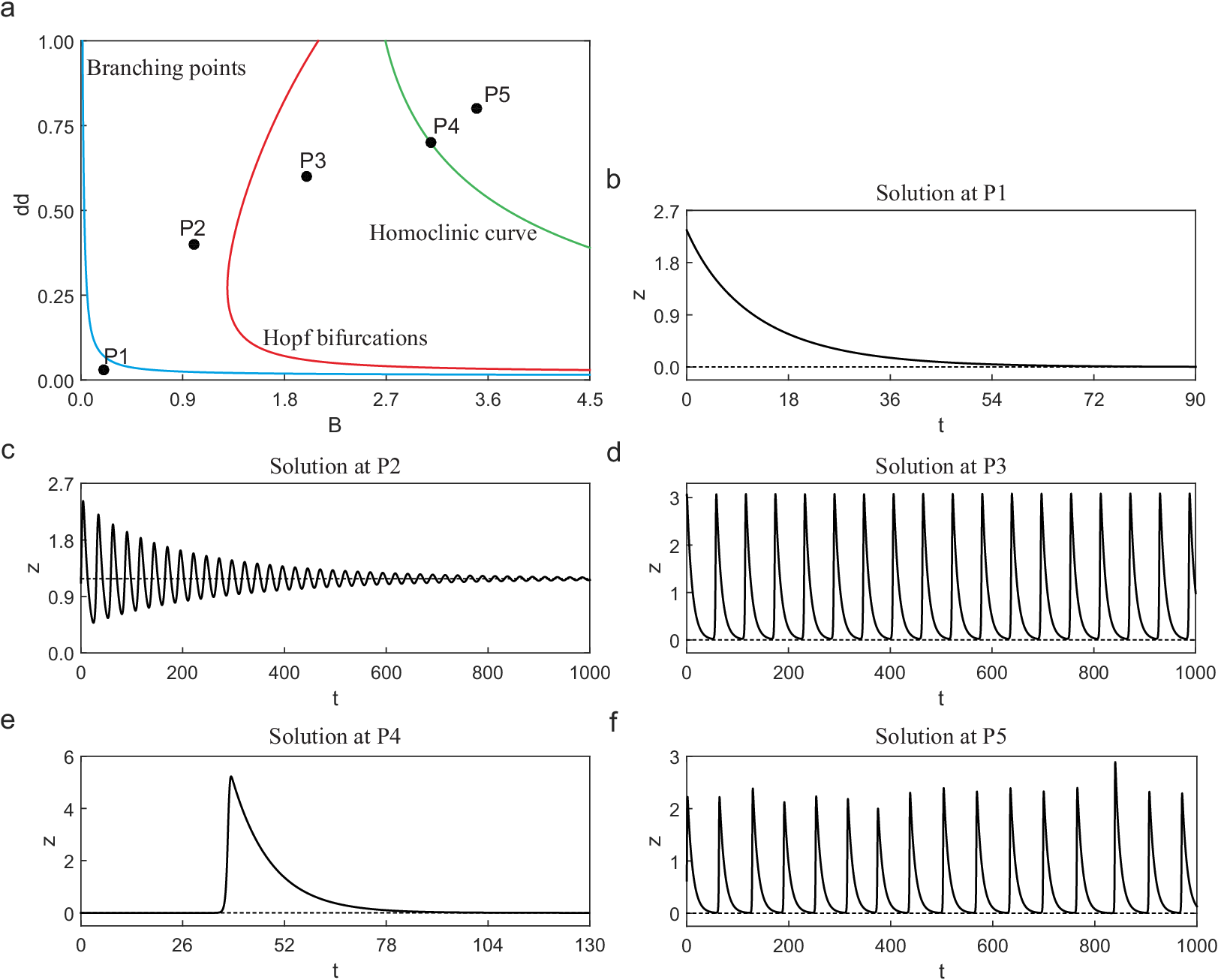}
\caption{(a) Two-parameter continuation with respect to $\beta$ and
$\delta$ of the codimension-1 bifurcations BP1 (in blue) and H1 (in
red, see Fig.\ \ref{Figure1}(a)), together with the
homoclinic curve computed in Fig.\ \ref{Figure3} (in
green). Panels (b)--(f) show time plots corresponding to the test
points P1 ($\beta=0.2$, $\delta=0.03$), P2 ($\beta=1$,
$\delta=0.4$), P3 ($\beta=2$, $\delta=0.6$), P4 ($\beta=3.0918$,
$\delta=0.7$) and P5 ($\beta=3.5$,
$\delta=0.8$).}\label{Figure4}
\end{figure}

Our numerical investigation has revealed the presence of two local
and one global bifurcation in the predator-prey
model~\eqref{eq:scaled}, which are responsible for significant
changes in the system response under parameter variations. In our
study, we are considering the predator's features (search rate and
assimilation efficiency) as the parameters of interest and we will
investigate how the system dynamics can be classified according to
those features. For this purpose, we will use COCO's capability to
carry out the numerical continuation of codimension-1 bifurcations
when two parameters are varied simultaneously. Specifically, we will
perform the numerical continuation of the bifurcation points H1 and
BP1 found in Fig.\ \ref{Figure1}(a) and the homoclinic
bifurcation, with respect to the parameters of interest $\beta$ and
$\delta$. The result can be seen in Fig.\
\ref{Figure4}(a). This figure presents three curves
corresponding to the continuation of the bifurcation points
mentioned before. The lowest curve (in blue) represents
$(\beta,\delta)$-tuples producing  branching points. In this way,
the curve divides locally the parameter space into two regions: one
for which the predator is present in the ecosystem, and one where
the predator population disappears. The curve indicates that to
ensure the predator's survival, a low assimilation efficiency has to
be compensated with a high search rate and vice versa. A second
curve (plotted in red) represents the numerical continuation of the
Hopf point H1 mentioned before. This curve then defines a boundary
for the appearance of endemic periodic solutions, where all
populations coexist in the ecosystem following an oscillatory
pattern. Finally, the third curve (in green) represents the
numerical continuation of near-homoclinic orbits described
previously, which gives an approximation of a curve of homoclinic
bifurcations. Crossing this curve from below destroys periodic
solutions via this global bifurcation, while all equilibria of the
system (extinction, predator extinction and coexistence) are
unstable. Further numerical investigation reveals the presence of
quasi-periodic orbits, as can be seen in Fig.\
\ref{Figure4}(f). In this figure, panels (b)--(f)
show time profiles corresponding to an excursion across the
two-parameter bifurcation diagram in panel (a).

\begin{figure}[H]
\centering \psfrag{a}{\large(a)}\psfrag{b}{\large(b)}
\psfrag{r}{\Large$r$}\psfrag{K}{\Large$K$}\psfrag{p}{\Large$\rho$}
\includegraphics[width=\textwidth]{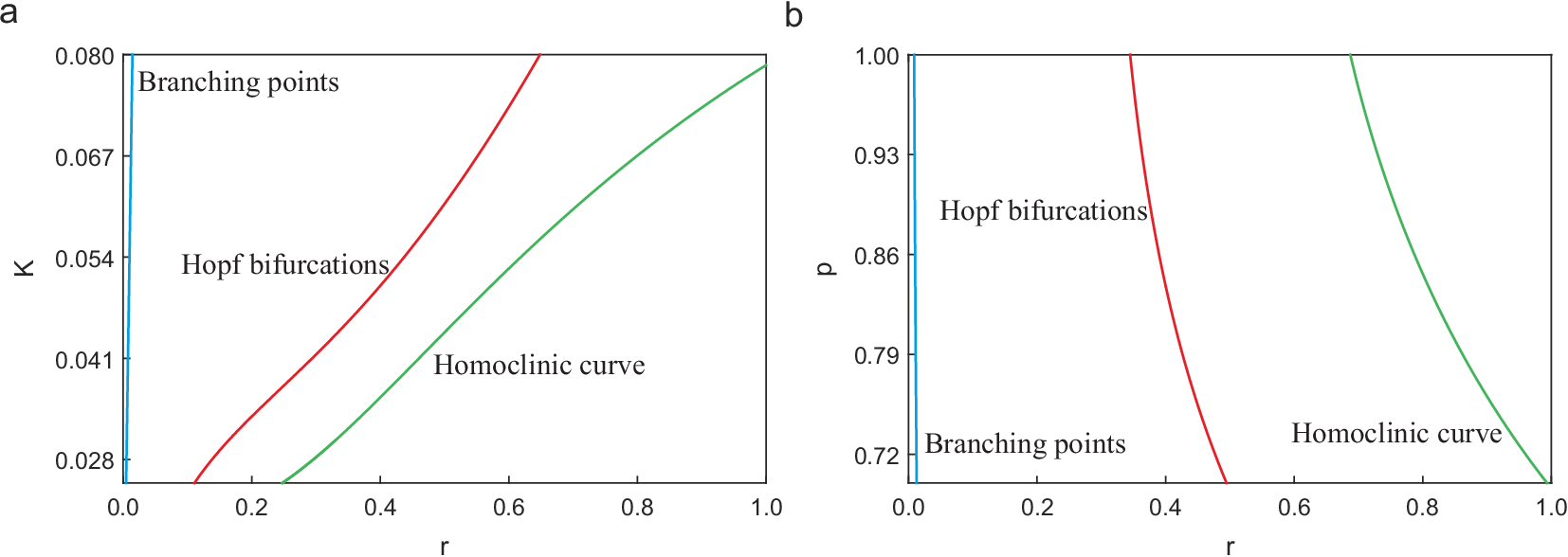}
\caption{Two-parameter continuation of branching points (in blue),
Hopf bifurcations (in red) and homoclinic orbits (in green), with
respect to $(r,K)$ (panel (a)) and $(r,\rho)$ (panel (b)), computed
for the parameter values given in Table~\ref{tab:cont}. The color
code is as in Fig.\
\ref{Figure4}(a).}\label{Figure5}
\end{figure}

To conclude our investigation, we will now consider the system
dynamics under parameter variations related to the prey's
characteristics. Specifically, we will carry out a two-parameter
continuation of the codimension-1 bifurcations found before
(branching point, Hopf, homoclinic) with respect to the prey's
intrinsic growth rate $r$ and the competition coefficient $K$. The
result can be seen in Fig.\ \ref{Figure5}(a). Similar to
Fig.\ \ref{Figure4}(a), this diagram shows three
curves representing $(r,K)$-values producing the codimension-1
bifurcations specified above. The resulting curves reveal that in
the considered biological scenario, the predator population enjoys
strong survival possibilities, as the region for which the predators
disappear is very narrow (the area between the vertical axis and the
blue curve corresponding to branching points). This means that only
significantly low prey's intrinsic growth rates would endanger the
predator sustainability in the system. A similar scenario is
encountered if we now take the prey's intrinsic growth rate $r$ and
the predation reduction coefficient $\rho$, see Fig.\
\ref{Figure5}(b). As before, the curves depicted in this
diagram allows the classification of the model behavior according to
the considered parameters corresponding to the prey's features.

\section{Discussion and concluding remarks}

In the present work, we propose a three-stage, two-species
predator-prey model that incorporates the role of honest signals and
cues. We subdivide the prey population into the subpopulation of
inexperienced individuals that are not able to transmit honest
signals (typically infants and juveniles) and that of experienced
individuals that are able to do so. Unlike the widely-used
game-theoretic models, our model focuses on the effect of honest
signals, cues and the associated learning among inexperienced
individuals to shape their genetically-inherited anti-predation
technique on the population level. The standard birth term, logistic
competition term, and predation term inspired by Holling's time
budgeting principle are involved in the model. A border line between
honest signal and cue is controlled by the parameter $\rho$. The
value of $\rho$ close to 0 indicates that the average experienced
prey individual successfully transmit signals and they are digested
as honest signals by all the nearby predators, such that no catch
will be pursued against it. This condition emphasizes the presence
of what we call error-free honest signals. When $\rho$ is close to
1, then some predators accept the transmitted signals as honest, but
some others do not (maybe those in distant ranges) and such flawed
honest signals even help the latter locate the whereabout of the
prey individual. The flawed honest signals, when benefitting the
predators exclusively, turn into cues. We further model a learning
rate for the inexperienced prey conveying the information regarding
different learning speeds when the predator density is around a
certain threshold $\eta$. Below the threshold, the learning rate
(speed) is relatively small (smaller than one) and decreases only
slowly as the experienced prey tends to be more abundant, that is,
due to overwhelming parental protection. Above the threshold, the
learning rate is relatively large and increases slowly with the
experienced prey density, also due to parental protection. For the
sake of simplification, we further performed a scaling of the
original model with eleven parameters to obtain that with seven
parameters. Buckingham $\pi$--Theorem suggests that the eleven
parameters can be cut to the least of eight parameters using
dimensional analysis, otherwise biological questions arise. In our
case, imposing $\sqrt{\alpha}=\eta$ in the learning rate due to the
non-observability of $\alpha$ is a stringent assumption. Even though
the case might be representable by a certain prey type in some
approximation, finding the exact types that fall into the equality
can however be a hard task. Therefore, what we draw from the
numerical bifurcation analysis are worth of consideration that they
are applied only to this restrictive case.

We point out several biological remarks regarding the numerical
bifurcation analysis. We argue from the model that a certain value
for the predator's search rate $\beta$ sets a threshold or the lower
bound for the coexistence between predator and prey population. The
asymptotic predator density increases for larger $\beta$ and
decreases as $\beta$ increases further, indicating a trade-off
between slow uptake, which leaves more food for population
sustenance, and fast uptake, which induces food's scarcity. When
$\beta$ tends to be larger, the predator population exhibits
transient cycles between persistence and near-extinction,
represented by the bifurcating stable limit cycles from a Hopf
point. Apparently, larger values of $\beta$ depict typical (bounded)
cycles with the period tending to blow up where the cycle ultimately
collides with the saddle-node extinction equilibrium. At this stage,
the predator and prey population are attracted to a homoclinic orbit
and ultimately go to extinction after long time. The same algorithm
can be used to reproduce the $(\beta,\delta)$-values at which
homoclinic orbits with the extinction equilibrium as the homoclinic
point occur. Our computation reveals that homoclinic orbits can be
seen through relatively large values of $\beta$ and the predator's
assimilation efficiency $\delta$, with a conflicting mode. That is,
they occur specifically when $\beta$ and $\delta$ cannot be
independently large. Any value exceeding the tuple returns a
quasi-periodic orbit where the dynamics of the predator and prey
population remains cyclical. All the preceding expositions indicate
the exclusiveness of the homoclinic orbits, which encounter in
between cyclical dynamics under a specific set of
$(\beta,\delta)$-values. A similar situation as for $\beta$ also
happens with $\delta$. A certain threshold is needed for the
long-term sustenance of the predator population. Then, the
asymptotic predator density increases and decreases with $\delta$,
which shows the impact of slow and fast reproduction on the scarcity
of prey. With larger values, similar transient cycles between
persistence and near-extinction occur, but they fold up into
asymptotic densities as $\delta$ increases further. This nontrivial
finding indicates that a stable asymptotic density of the predator
individuals is expected when they produce offspring much smaller
than or nearly of the same magnitude as the captured prey
individuals per unit area. As far as the parameters $K,r,\rho$ are
concerned, similar arguments as before can be drawn. We found that
reducing the competition coefficient $K$ under a moderate intrinsic
growth rate $r$ (around 0.5) gives an increasing degree of
complexity in the asymptotic behavior of the predator--prey
dynamics. This is to say that increasing the prey carrying capacity
(due to $C\sim 1\slash K$) not only accommodates prey with enough
space and food resources, but also facilitates the discovery of
nontrivial trajectories: transient cycles, homoclinic orbits and
quasi-periodic orbits. Unlike the $(\beta,\delta)$-tuple, increasing
$K$ and $r$ simultaneously can preserve the type of orbit
discovered. Further numerical finding reveals that the separator
between honest signal and cue $\rho$ is only sensitive if the
intrinsic growth rate $r$ in relatively large (above 0.6). A small
$\rho$ can only lead the model solution to the stable asymptotic
predator--prey density, meaning that when the experienced prey
always bears error-free honest signals, the entire populations
coexist and verge on an equilibrium after long time. When the
quality of the signals is compromised by a certain fraction of
hunting predators ($\rho$ increases), the equilibrium exhibits
instability, yet the stability is taken over by the bifurcating
limit cycles, inducing unpredictability of the dynamics that runs
around persistence and near-extinction up to a certain degree. Until
then, quasi-periodic orbits dominate at sufficiently large $\rho$
(close to 1), which add more degree of unpredictability. We thus
argue from the model that error-free honest signals lead to not only
a stable predator-prey dynamics but also predictable behavior.

\section*{Acknowledgements}

The second author has been supported by the DAAD Visiting
Professorships programme at the University of Koblenz-Landau,
Germany.

\bibliography{bibli}

\begin{thebibliography}{10}

\bibitem{owren2010redefining}
M.~J. Owren, D.~Rendall, and M.~J. Ryan, ``Redefining animal signaling:
  influence versus information in communication,'' {\em Biology \& Philosophy},
  vol.~25, no.~5, pp.~755--780, 2010.

\bibitem{smith1997behavior}
W.~J. Smith, ``The behavior of communicating, after twenty years,'' in {\em
  Perspectives in Ethology} (D.~H. Owings, M.~D. Beecher, and N.~S. Thompson,
  eds.), Boston, MA: Springer US, 1997.

\bibitem{hauser1996evolution}
M.~D. Hauser, {\em The evolution of communication}.
\newblock Cambridge, MA: MIT press, 1996.

\bibitem{bugnyar2001food}
T.~Bugnyar, M.~Kijne, and K.~Kotrschal, ``Food calling in ravens: are yells
  referential signals?,'' {\em Animal Behaviour}, vol.~61, no.~5, pp.~949--958,
  2001.

\bibitem{lehmann2014cues}
K.~D. Lehmann, B.~W. Goldman, I.~Dworkin, D.~M. Bryson, and A.~P. Wagner,
  ``From cues to signals: evolution of interspecific communication via
  aposematism and mimicry in a predator-prey system,'' {\em PloS one}, vol.~9,
  no.~3, p.~e91783, 2014.

\bibitem{johnstone1992continuous}
R.~A. Johnstone and A.~Grafen, ``The continuous sir philip sidney game: a
  simple model of biological signalling,'' {\em Journal of theoretical
  Biology}, vol.~156, no.~2, pp.~215--234, 1992.

\bibitem{leal1999honest}
M.~Leal, ``Honest signalling during prey--predator interactions in the lizard
  anolis cristatellus,'' {\em Animal Behaviour}, vol.~58, no.~3, pp.~521--526,
  1999.

\bibitem{zahavi1975mate}
A.~Zahavi, ``Mate selection a selection for a handicap,'' {\em Journal of
  theoretical Biology}, vol.~53, no.~1, pp.~205--214, 1975.

\bibitem{zahavi1977cost}
A.~Zahavi, ``The cost of honesty (further remarks on the handicap
  principle).,'' {\em Journal of theoretical Biology}, vol.~67, no.~3,
  pp.~603--605, 1977.

\bibitem{zahavi1987theory}
A.~ZAHAVI, ``The theory of signal selection and some of its implications,''
  {\em International Symposium of Biological Evolution, 1987}, pp.~pp 305--327,
  1987.

\bibitem{dawkins1978animal}
R.~Dawkins and J.~R. Krebs, ``Animal signals: information or manipulation,''
  {\em Behavioural ecology: An evolutionary approach}, vol.~2, pp.~282--309,
  1978.

\bibitem{krebs1984behavioral}
J.~Krebs, R.~Dawkins, and N.~Davies, ``Behavioral ecology: An integrated
  approach,'' 1984.

\bibitem{zahavi1999handicap}
A.~Zahavi and A.~Zahavi, {\em The handicap principle: A missing piece of
  Darwin's puzzle}.
\newblock Oxford University Press, Oxford, UK, 1999.

\bibitem{grafen1990biological}
A.~Grafen, ``Biological signals as handicaps,'' {\em Journal of theoretical
  biology}, vol.~144, no.~4, pp.~517--546, 1990.

\bibitem{szamado2015does}
S.~Sz{\'a}mad{\'o} and D.~J. Penn, ``Why does costly signalling evolve?
  challenges with testing the handicap hypothesis,'' {\em Animal behaviour},
  vol.~110, p.~e9, 2015.

\bibitem{bergstrom2002separating}
C.~T. Bergstrom, S.~Sz{\'a}mad{\'o}, and M.~Lachmann, ``Separating equilibria
  in continuous signalling games,'' {\em Philosophical Transactions of the
  Royal Society of London. Series B: Biological Sciences}, vol.~357, no.~1427,
  pp.~1595--1606, 2002.

\bibitem{getty1998handicap}
T.~Getty, ``Handicap signalling: when fecundity and viability do not add up,''
  {\em Animal Behaviour}, vol.~56, no.~1, pp.~127--130, 1998.

\bibitem{getty2006sexually}
T.~Getty, ``Sexually selected signals are not similar to sports handicaps,''
  {\em Trends in Ecology \& Evolution}, vol.~21, no.~2, pp.~83--88, 2006.

\bibitem{lachmann2001cost}
M.~Lachmann, S.~Szamado, and C.~T. Bergstrom, ``Cost and conflict in animal
  signals and human language,'' {\em Proceedings of the National Academy of
  Sciences}, vol.~98, no.~23, pp.~13189--13194, 2001.

\bibitem{szamado2011cost}
S.~Sz{\'a}mad{\'o}, ``The cost of honesty and the fallacy of the handicap
  principle,'' {\em Animal Behaviour}, vol.~81, no.~1, pp.~3--10, 2011.

\bibitem{polnaszek2014not}
T.~J. Polnaszek and D.~W. Stephens, ``Why not lie? costs enforce honesty in an
  experimental signalling game,'' {\em Proceedings of the Royal Society B:
  Biological Sciences}, vol.~281, no.~1774, p.~20132457, 2014.

\bibitem{clifton2016handicap}
S.~M. Clifton, R.~I. Braun, and D.~M. Abrams, ``Handicap principle implies
  emergence of dimorphic ornaments,'' {\em Proceedings of the Royal Society B:
  Biological Sciences}, vol.~283, no.~1843, p.~20161970, 2016.

\bibitem{hasson1997towards}
O.~Hasson, ``Towards a general theory of biological signaling,'' {\em Journal
  of Theoretical Biology}, vol.~185, no.~2, pp.~139--156, 1997.

\bibitem{johnstone1992error}
R.~A. Johnstone and A.~Grafen, ``Error-prone signalling,'' {\em Proceedings of
  the Royal Society of London. Series B: Biological Sciences}, vol.~248,
  no.~1323, pp.~229--233, 1992.

\bibitem{johnstone1994honest}
R.~A. Johnstone, ``Honest signalling, perceptual error and the evolution of
  `all-or-nothing'displays,'' {\em Proceedings of the Royal Society of London.
  Series B: Biological Sciences}, vol.~256, no.~1346, pp.~169--175, 1994.

\bibitem{smith2003animal}
J.~M. Smith and D.~Harper, {\em Animal signals}.
\newblock Oxford University Press, Oxford, UK, 2003.

\bibitem{laidre2013animal}
M.~E. Laidre and R.~A. Johnstone, ``Animal signals,'' {\em Current Biology},
  vol.~23, no.~18, pp.~R829--R833, 2013.

\bibitem{rendall2009animal}
D.~Rendall, M.~J. Owren, and M.~J. Ryan, ``What do animal signals mean?,'' {\em
  Animal Behaviour}, vol.~78, no.~2, pp.~233--240, 2009.

\bibitem{holling1959components}
C.~S. Holling, ``The components of predation as revealed by a study of
  small-mammal predation of the european pine sawfly,'' {\em The Canadian
  Entomologist}, vol.~91, no.~5, pp.~293--320, 1959.

\bibitem{jeschke2002predator}
J.~M. Jeschke, M.~Kopp, and R.~Tollrian, ``Predator functional responses:
  discriminating between handling and digesting prey,'' {\em Ecological
  Monographs}, vol.~72, no.~1, pp.~95--112, 2002.

\bibitem{smith1995animal}
M.~J. Smith and D.~G. Harper, ``Animal signals: models and terminology,'' {\em
  Journal of theoretical biology}, vol.~177, no.~3, pp.~305--311, 1995.

\bibitem{FF1988}
C.~D. FitzGibbon and J.~H. Fanshawe, ``Stotting in thomson's gazelles: an
  honest signal of condition,'' {\em Behavioral Ecology and Sociobiology},
  vol.~23, no.~2, pp.~69--74, 1988.

\bibitem{csanyi1993learning}
V.~Cs{\'a}nyi and A.~D{\'o}ka, ``Learning interactions between prey and
  predator fish,'' {\em Marine \& Freshwater Behaviour \& Phy}, vol.~23,
  no.~1-4, pp.~63--78, 1993.

\bibitem{brown2003social}
C.~Brown and K.~N. Laland, ``Social learning in fishes: a review,'' {\em Fish
  and fisheries}, vol.~4, no.~3, pp.~280--288, 2003.

\bibitem{boyd1988culture}
R.~Boyd and P.~J. Richerson, {\em Culture and the evolutionary process}.
\newblock University of Chicago press, Chicago USA, 1988.

\bibitem{Wein2008}
M.~Weinstock, ``The long-term behavioural consequences of prenatal stress,''
  {\em Neuroscience and Biobehavioral Reviews}, vol.~32, pp.~1073--1086, 2008.

\bibitem{Scho2011}
S.~J. Schoech, M.~A. Rensel, and R.~S. Heiss, ``Short and long-term effects of
  developmental corticosterone exposure on avian physiology, behavioral
  phenotype, cognition, and fitness: review,'' {\em Current Zoology}, vol.~57,
  no.~4, pp.~514--530, 2011.

\bibitem{Macca2014}
S.~Maccari, H.~J. Krugers, S.~Morley-Fletcher, M.~Szyf, and P.~J. Brunton,
  ``The consequences of early-life adversity: neurobiological, behavioural and
  epigenetic adaptations,'' {\em Journal of Neuroendocrinology}, vol.~26,
  pp.~707--723, 2014.

\bibitem{Crandall1980}
M.~G. Crandall and P.~H. Rabinowitz, ``Mathematical theory of bifurcation,'' in
  {\em Bifurcation Phenomena in Mathematical Physics and Related Topics}
  (C.~Bardos and D.~Bessis, eds.), (Dordrecht), pp.~3--46, Springer
  Netherlands, 1980.

\bibitem{Rabi1977}
P.~H. Rabinowitz, ``Some global results for nonlinear eigenvalue problems,''
  {\em Journal of Functional Analysis}, vol.~7, no.~3, pp.~487--513, 1971.

\bibitem{Golu1985}
M.~Golubitsky and D.~G. Schaeffer, {\em Singularities and Groups in Bifurcation
  Theory}, vol.~I.
\newblock New York: Springer, 1985.

\bibitem{Ma2005}
T.~Ma and S.~Wang, {\em Bifurcation Theory and Applications}, vol.~53 of {\em
  World Scientific Series on Nonlinear Science A}.
\newblock Singapore: World Scientific Publishing, 2005.

\bibitem{Krasno1984}
M.~Krasnoselskii and P.~Zabreiko, {\em Geometrical Methods of Nonlinear
  Analysis}.
\newblock New York: Springer, 1984.

\bibitem{Cushing1988}
J.~M. Cushing, {\em An Introduction to Structured Population Dynamics}.
\newblock CBMS-NSF Conference Series in Applied Mathematics, Philadelphia:
  Society for Industrial and Applied Mathematics, 1988.

\bibitem{Wijaya2017}
K.~P. Wijaya, Sutimin, E.~Soewono, and T.~G\"{o}tz, ``On the existence of a
  nontrivial equilibrium in relation to the basic reproductive number,'' {\em
  Journal of Applied Mathematics and Computer Science}, vol.~27, no.~3,
  pp.~623--636, 2017.

\bibitem{dankowicz2013}
H.~Dankowicz and F.~Schilder, {\em Recipes for continuation}.
\newblock Computational Science and Engineering, Philadelphia: SIAM, 2013.

\bibitem{auto97}
E.~J. Doedel, A.~R. Champneys, T.~F. Fairgrieve, Y.~A. Kuznetsov, B.~Sandstede,
  and X.-J. Wang, {\em {Auto97: Continuation and bifurcation software for
  ordinary differential equations (with HomCont)}}.
\newblock Computer Science, Concordia University, Montreal, Canada, 1997.
\newblock Available at \url{http://cmvl.cs.concordia.ca}.

\bibitem{matcont}
A.~Dhooge, W.~Govaerts, and Y.~A. Kuznetsov, ``{MATCONT: A MATLAB package for
  numerical bifurcation analysis of ODEs},'' {\em ACM Trans. Math. Software},
  vol.~29, no.~2, pp.~141--164, 2003.

\bibitem{guck}
J.~Guckenheimer and P.~Holmes, {\em Nonlinear Oscillations, Dynamical Systems,
  and Bifurcations of Vector Fields}, vol.~42 of {\em Applied Mathematical
  Sciences}.
\newblock New York: Springer-Verlag, 1993.
\newblock Fourth printing.

\bibitem{govaertslibro}
W.~Govaerts, {\em Numerical Methods for Bifurcations of Dynamical Equilibria}.
\newblock Philadelphia: SIAM, 2000.

\bibitem{velasco2000}
C.~M. Kribs-Zaleta and J.~X. Velasco-Hern\'andez, ``A simple vaccination model
  with multiple endemic states,'' {\em Mathematical Biosciences}, vol.~164,
  no.~2, pp.~183--201, 2000.

\bibitem{watmough2000}
P.~van~den Driessche and J.~Watmough, ``{A simple SIS epidemic model with a
  backward bifurcation},'' {\em J. Math. Biol.}, vol.~40, no.~6, pp.~525--540,
  2000.

\bibitem{liz2015}
M.~Liu, E.~Liz, and G.~R{\"o}st, ``{Endemic bubbles generated by delayed
  behavioral response: Global stability and bifurcation switches in an SIS
  model},'' {\em SIAM J. Appl. Math.}, vol.~75, no.~1, pp.~75--91, 2015.

\end{thebibliography}
\bibliographystyle{ieeetr}

\end{document}